\documentstyle[12pt,epsf]{article}
\begin{document}

\def\be{\begin{equation}}
\def\ee{\end{equation}}
\def\ba{\begin{eqnarray}}
\def\ea{\end{eqnarray}}

\title{Duality relations for $M$ coupled Potts models}
\author{Jesper Lykke Jacobsen \\
        LPTMS, b\^{a}timent~100,
        Universit\'e Paris-Sud, F-91405 Orsay, France}
\date{January 2000}
\maketitle

\begin{abstract}
 We establish explicit duality transformations for systems of $M$
 $q$-state Potts
 models coupled through their local energy density, generalising known
 results for $M=1,2,3$. The $M$-dimensional space of coupling constants
 contains a selfdual sub-manifold of dimension $D_M = [M/2]$.
 For the case $M=4$, the variation of the effective central charge
 along the selfdual surface is investigated by numerical transfer matrix 
 techniques. Evidence is given for the existence of a family of critical
 points, corresponding to conformal field theories with an extended $S_M$
 symmetry algebra.
\end{abstract}

For several decades, the $q$-state Potts model has been used to
model ferromagnetic materials \cite{Potts53}, and an impressive
number of results are known about it, especially in two dimensions
\cite{Wu82,Baxter82,Saleur91}. More recently, its random-bond counterpart
has attracted considerable attention \cite{Cardy99}, primarily because it
permits one to study how quenched randomness coupling
to the local energy density can modify the nature of a phase transition.

But despite the remarkable successes of conformal invariance applied to
pure two-dimensional systems, the amount of analytical results on the
random-bond Potts model is rather scarce. Usually the disorder is dealt
with by introducing $M$ replicas of the original model, with mutual
energy-energy interactions, and taking the limit $M \to 0$. The price to
be paid is however that the resulting system loses many of the properties
(such as unitarity) that lie at the heart of conventional conformal
field theory \cite{Ludwig87,Ludwig90}.

Very recently, an alternative approach was suggested by Dotsenko {\em et al}
\cite{Dotsenko}. These authors point out that the perturbative
renormalisation group \cite{Ludwig87} (effectively an expansion around the
Ising model in the small parameter $\varepsilon = q-2$) predicts the
existence of a non-trivial infrared fixed point at interlayer coupling
$g_* \propto -\varepsilon/(M-2) + {\cal O}(\varepsilon^2)$,
so that the regions $M<2$ and $M>2$ are somehow dual upon changing the
sign of the coupling constant%
\footnote{The case $M=2$ is special: For $q=2$ (the Ashkin-Teller model)
the coupling presents a marginal perturbation, giving rise to a halfline
of critical points along which the critical exponents vary
continuously \cite{Baxter82}. On the other hand, for $q>2$ where the
perturbation is relevant, the model is still integrable, but now presents
a mass generation leading to non-critical behaviour \cite{Vaysburd}.}.
More interestingly, for $M=3$ they identify
the exact lattice realisation of a critical theory with exponents
consistent with those of the perturbative treatment, and they conjecture
that this generalises to any integer $M \ge 3$. Their proposal is then to
study this class of coupled models, which are now unitary by definition,
and only take the limit $M \to 0$ once the exact expressions for the
various critical exponents have been worked out. One could hope to attack
this task by means of extended conformal field theory, thus combining the
$Z_q$ symmetry of the spin variable by a non-abelian $S_M$ symmetry upon
permuting the replicas.

Clearly, a first step in this direction is to identify the lattice models
corresponding to this series of critical theories, parametrised by the
integer $M \ge 3$. For $M=3$ this was achieved \cite{Dotsenko} by working
out the duality relations for $M$ coupled Potts models on the square lattice,
within the $M$-dimensional space of coupling constants giving rise to
$S_M$ symmetric interactions amongst the lattice energy operators of the
replicas. Studying numerically the variation of the effective central
charge \cite{Zamolodchikov} along the resulting selfdual line, using a novel
and very powerful transfer matrix technique, the critical point was
unambiguously identified with one of the endpoints of that line.

Unfortunately it was hard to see how such duality relations could be
extended to the case of general $M$. The calculations in Ref.~\cite{Dotsenko}
relied on a particular version \cite{Wu} of the method of lattice Fourier
transforms \cite{Savit}, already employed for $M=2$ two decades
ago \cite{Domany}. Though perfectly adapted to the case of linear
combinations of cosinoidal interactions within a single (vector) Potts model
\cite{Savit}, this approach led to increasingly complicated algebra
when several coupled models were considered. Moreover, it seemed
impossible to recast the end results in a reasonably simple form
for larger $M$.

In the present publication we wish to assess whether such a scenario of a
unique critical point with an extended $S_M$ symmetry can indeed be expected
to persist in the general case of $M \ge 3$ symmetrically coupled models.
We explicitly work out the duality transformations for any $M$, and
show that they can be stated in a very simple form [Eq.~(\ref{dual})]
after redefining the coupling constants.

The lattice identification of the $M=3$ critical point
in Ref.~\cite{Dotsenko} crucially relied on the existence of a
{\em one-parameter} selfdual manifold, permitting only two possible
directions of the initial flow away from the decoupling fixed point.
We find in general a richer structure with an $[M/2]$-dimensional selfdual
manifold. Nonetheless, from a numerical study of the case $M=4$ we end up
concluding that the uniqueness of the non-trivial fixed point can be expected
to persist, since the decoupling fixed point acts as a saddlepoint of the
effective central charge.

Consider then a system of $M$ identical planar lattices, stacked on top of one
another. On each lattice site $i$, and for each layer $\mu = 1,2,\ldots,M$,
we define a Potts spin $\sigma^{(\mu)}_i$
that can be in any of $q = 2,3,\ldots$ distinct states.
The layers interact by means of the reduced hamiltonian
\be
 {\cal H} = \sum_{\langle ij \rangle} {\cal H}_{ij},
\ee
where $\langle ij \rangle$ denotes the set of lattice edges, and an
$S_M$ symmetric nearest-neighbour interaction is defined as
\be
 {\cal H}_{ij} = - \sum_{m=1}^M K_m
   \sum_{\mu_1 \neq \mu_2 \neq \cdots \mu_m}'
   \prod_{l=1}^m \delta \left( \sigma^{(\mu_l)}_i,\sigma^{(\mu_l)}_j \right).
\ee
By definition the primed summation runs over the ${M \choose m}$ terms
for which the indices $1 \le \mu_l \le M$ with $l=1,2,\ldots,m$ are all
different, and $\delta(x,y) = 1$ if $x=y$ and zero otherwise.

For $M=1$ the model thus defined reduces to the conventional Potts model,
whilst for $M=2$ it is identical to the Ashkin-Teller like model
considered in Ref.~\cite{Domany}, where the Potts models of either layer are
coupled through their local energy density. For $M>2$, additional multi-energy
interactions between several layers have been added, since such interactions
are generated by the duality transformations, as we shall soon see.
However, from the point of view of conformal field theory these supplementary
interactions are irrelevant in the continuum limit.
The case $M=3$ was discussed in Ref.~\cite{Dotsenko}.

By means of a generalised Kasteleyn-Fortuin transformation \cite{Kasteleyn}
the local Boltzmann weights can be recast as
\be
 \exp(-{\cal H}_{ij}) = \prod_{m=1}^M
   \prod_{\mu_1 \neq \mu_2 \neq \cdots \mu_m}'
   \left[1 + \left( {\rm e}^{K_m}-1 \right)
   \prod_{l=1}^m \delta(\sigma^{(\mu_l)}_i,\sigma^{(\mu_l)}_j) \right].
 \label{Boltzmann}
\ee

In analogy with the case of $M=1$, the products can now be expanded so as
to transform the original Potts model into its associated random cluster
model. To this end we note that Eq.~(\ref{Boltzmann}) can be rewritten
in the form
\be
 \exp(-{\cal H}_{ij}) = b_0 + \sum_{m=1}^M b_m
   \sum_{\mu_1 \neq \mu_2 \neq \cdots \mu_m}'
   \prod_{l=1}^m \delta(\sigma^{(\mu_l)}_i,\sigma^{(\mu_l)}_j),
 \label{Boltzmann1}
\ee
defining the coefficients $\{ b_m \}_{m=0}^M$.
The latter can be related to the physical coupling constants
$\{ K_m \}_{m=1}^M$ by evaluating Eqs.~(\ref{Boltzmann}) and
(\ref{Boltzmann1}) in the situation where precisely $m$ out of the
$M$ distinct Kronecker $\delta$-functions are non-zero.
Clearly, in this case Eq.~(\ref{Boltzmann}) is equal to
${\rm e}^{J_m}$, where 
\be
 J_m = \sum_{k=1}^m {m \choose k} K_k
 \label{bond1}
\ee
for $m \ge 1$, and we set $J_0 = K_0 = 0$.
On the other hand, we find from Eq.~(\ref{Boltzmann1})
that this must be equated to
$\sum_{k=0}^m {m \choose k} b_k$.
This set of $M+1$ equations can be solved for the $b_k$ by recursion,
considering in turn the cases $m=0,1,\ldots,M$. After some algebra,
the edge weights $b_k$ (for $k \ge 0$) are then found as
\be
 b_k = \sum_{m=0}^k (-1)^{m+k} {k \choose m} {\rm e}^{J_m}.
 \label{bond2}
\ee

The partition function in the spin representation
\be
 Z = \sum_{\lbrace \sigma \rbrace}
     \prod_{\langle ij \rangle} \exp(-{\cal H}_{ij})
\ee
can now be transformed into the random cluster representation as
follows. First, insert Eq.~(\ref{Boltzmann1}) on the right-hand side
of the above equation, and imagine expanding the product over the
lattice edges $\langle ij \rangle$. To each term in the resulting sum
we associate an edge colouring ${\cal G}$ of the $M$-fold replicated
lattice, where an edge $(ij)$ in layer $m$ is considered to be
coloured (occupied) if the term contains the factor
$\delta(\sigma^{(m)}_i,\sigma^{(m)}_j)$, and uncoloured (empty) if it does not.
[In this language, the couplings $J_k$ correspond to the local energy density
summed over all possible permutations of precisely $k$ simultaneously
coloured edges.]

The summation over the spin variables $\{ \sigma \}$ is now trivially
performed, yielding a factor of $q$ for each connected component
(cluster) in the colouring graph. Keeping track of the prefactors
multiplying the $\delta$-functions, using Eq.~(\ref{Boltzmann1}), we
conclude that
\be
 Z = % \sum_{\lbrace \sigma \rbrace} {\rm e}^{-{\cal H}} =
   \sum_{\cal G} \prod_{m=1}^M q^{C_m} b_m^{B_m},
 \label{Z-cluster}
\ee
where $C_m$ is the number of clusters in the $m$th layer, and $B_m$ is the
number of occurencies in ${\cal G}$ of a situation where precisely
$m$ ($0 \le m \le M$) edges placed on top of one another have been
simultaneously coloured.

It is worth noticing that the random cluster description of the model
has the advantage that $q$ only enters as a parameter. By analytic
continuation one can thus give meaning to a non-integer number of states.
The price to be paid is that the $C_m$ are, a priori, non-local quantities.

In terms of the edge variables $b_m$ the duality transformation of the
partition function is easily worked out. For simplicity we shall
assume that the couplings constants $\{ K_m \}$ are identical between
all nearest-neighbour pairs of spins, the generalisation to an
arbitrary inhomogeneous distribution of couplings being trivial.
By analogy with the case $M=1$, a given colouring configuration
${\cal G}$ is taken to be dual to a colouring configuration
$\tilde{\cal G}$ of the dual lattice obtained by applying the following
duality rule: {\em Each coloured edge intersects an uncoloured dual
edge}, and vice versa. In particular, the demand that the
configuration ${\cal G}_{\rm full}$ with all lattice edges coloured be
dual to the configuration ${\cal G}_{\rm empty}$ with no coloured
(dual) edge fixes the constant entering the duality transformation.
Indeed, from Eq.~(\ref{Z-cluster}), we find that ${\cal G}_{\rm full}$ has
weight $q^M b_M^E$, where $E$ is the total number of lattice edges,
and ${\cal G}_{\rm empty}$ is weighted by $q^{MF} \tilde{b}_0^E$, where $F$ is
the number of faces, including the exterior one.
We thus seek for a duality transformation of the form
$q^{MF} \tilde{b}_0^E Z(\{ b_m \}) = q^M b_M^E \tilde{Z}(\{ \tilde{b}_m \})$,
where for any configuration ${\cal G}$ the edge weights must transform
so as to keep the same relative weight between ${\cal G}$ and
${\cal G}_{\rm full}$ as between $\tilde{\cal G}$ and ${\cal G}_{\rm empty}$.

An arbitrary colouring configuration ${\cal G}$ entering
Eq.~(\ref{Z-cluster}) can be generated by applying a finite number of
changes to ${\cal G}_{\rm full}$, in which an edge of weight $b_M$ is
changed into an edge of weight $b_m$ for some $m=0,1,\ldots,M-1$.
By such a change, in general, a number $k \le M-m$ of pivotal bonds are
removed from the colouring graph, thus creating $k$ new clusters, and
the weight relative to that of ${\cal G}_{\rm full}$ will change by
$q^k b_m / b_M$. On the other hand, in the dual configuration
$\tilde{\cal G}$ a number $M-m-k$ of clusters will be lost, since each
of the $k$ new clusters mentioned above will be accompanied by the
formation of a {\em loop} in $\tilde{\cal G}$. The weight change relative to
${\cal G}_{\rm empty}$ therefore amounts to
$\tilde{b}_{M-m}/(\tilde{b}_0 q^{M-m-k})$.
Comparing these two changes we see that the factors of $q^k$ cancel
nicely, and after a change of variables $m \to M-m$
the duality transformation takes the simple form
\be
 \tilde{b}_m = \frac{q^m b_{M-m}}{b_M} \ \ \ \ \mbox{ for } m=0,1,\ldots,M,
 \label{dual}
\ee
the relation with $m=0$ being trivial.

Selfdual solutions can be found by imposing $\tilde{b}_m = b_m$.
However, this gives rise to only $\left[ \frac{M+1}{2} \right]$
independent equations
\be
 b_{M-m} = q^{M/2-m} b_m \ \ \ \ \mbox{ for }
                                   m=0,1,\ldots,\left[\frac{M-1}{2}\right],
 \label{selfdual}
\ee
and the $M$-dimensional parameter space $\{ b_m \}_{m=1}^M$,
or $\{ K_m \}_{m=1}^M$, thus has a selfdual sub-manifold of
dimension $D_M = \left[ \frac{M}{2} \right]$.
In particular, the ordinary Potts model ($M=1$) has a unique selfdual point,
whilst for $M=2$ \cite{Domany} and $M=3$ \cite{Dotsenko} one has a line
of selfdual solutions.

Our main result is constituted by Eqs.~(\ref{bond1}) and (\ref{bond2}) relating
the physical coupling constants $\{ K_m \}$ to the edge weights $\{ b_m \}$,
in conjunction with Eqs.~(\ref{dual}) and (\ref{selfdual}) giving the explicit
(self)duality relations in terms of the latter.

Since the interaction energies entering Eq.~(\ref{Boltzmann}) are invariant
under a simultaneous shift of all Potts spins, an alternative way of
establishing the duality transformations procedes by Fourier
transformation of the energy gaps \cite{Wu}.
This method was used in Refs.~\cite{Domany} and
\cite{Dotsenko} to work out the cases $M=2$ and $M=3$ respectively.
However, as $M$ increases this procedure very quickly becomes quite involved.
To better appreciate the ease of the present approach, let
us briefly pause to see how the parametrisations of the selfdual lines
for $M=2,3$, expressed in terms of the couplings $\{ K_m \}$, 
can be reproduced in a most expedient manner.

For $M=2$, Eq.~(\ref{selfdual}) gives $b_2 = q$, where from
Eqs.~(\ref{bond1}) and (\ref{bond2})
$b_2 = {\rm e}^{2K_1 + K_2} - 2 {\rm e}^{K_1} + 1$.
Thus
\be
  {\rm e}^{K_2} = \frac{2 {\rm e}^{K_1} + (q-1)}{{\rm e}^{2 K_1}},
\ee
in accordance with Ref.~\cite{Domany}.
Similarly, for $M=3$ one has $b_1 = q b_2 / b_3 = b_2 / \sqrt{q}$ with
$b_1 = {\rm e}^{K_1} - 1$, $b_2$ as before,
and $b_3 = {\rm e}^{3K_1+3K_2+K_3}-3{\rm e}^{2K_1+K_2}+3{\rm e}^{K_1}-1$.
This immediately leads to the result given in Ref.~\cite{Dotsenko}:
\ba
  {\rm e}^{K_2} &=& \frac{(2+\sqrt{q}){\rm e}^{K_1} - (1+\sqrt{q})}
                         {{\rm e}^{2K_1}}, \\
  {\rm e}^{K_3} &=& \frac{3({\rm e}^{K_1}-1)(1+\sqrt{q}) + q^{3/2} + 1}
                         {\left[ (2+\sqrt{q}){\rm e}^{K_1}-(1+\sqrt{q})
                          \right]^3} \, {\rm e}^{3 K_1}. \nonumber
\ea

Returning now to the general case, we notice that the selfdual manifold
always contains two special points for which the behaviour of the $M$
coupled models can be related to that of a single Potts model.
At the first such point,
\be
 b_m = q^{m/2} \ \ \ \ \mbox{ for } m=0,1,\ldots,\left[\frac{M}{2}\right],
\ee
one has $K_1 = \log(1+\sqrt{q})$ and $K_m = 0$ for $m=2,3,\ldots,M$,
whence the $M$ models simply decouple.
The other point
\be
 b_m = \delta(m,0) \ \ \ \ \mbox{ for } m=0,1,\ldots,\left[\frac{M}{2}\right]
\ee
corresponds to $K_m = 0$ for $m=1,2,\ldots,M-1$ and
$K_M = \log(1+q^{M/2})$, whence the resulting model is equivalent to
a single $q^M$-state Potts model.
Evidently, for $M=1$ these two special points coincide.

Specialising now to the case of a regular two-dimensional lattice, it
is well-known that at the two special points the model undergoes a phase
transition, which is continuous if the effective number of states
($q$ or $q^M$ as the case may be) is $\le 4$ \cite{Baxter73}.
In Ref.~\cite{Dotsenko} the question was raised whether one in general
can identify further non-trivial critical theories on the selfdual
manifolds. In particular it was argued that for $M=3$ there is indeed
such a point, supposedly corresponding to a conformal field theory with
an extended $S_3$ symmetry.

To get an indication whether such results can be expected to generalise
also to higher values of $M$, we have numerically computed the effective
central charge of $M=4$ coupled models along the two-dimensional
selfdual surface. We were able to diagonalise the transfer matrix for
strips of width $L=4,6,8$ lattice constants in the equivalent loop model.
Technical details of the simulations have been reported in
Ref.~\cite{Dotsenko}. Relating the specific free energy $f_0(L)$
to the leading eigenvalue of the transfer matrix in the standard way,
two estimates of the effective central charge, $c(4,6)$ and $c(6,8)$, 
were then obtained by fitting data for two
consecutive strip widths according to \cite{Cardy86}
\be
  f_0(L) = f_0(\infty) - \frac{\pi c}{6 L^2} + \cdots.
  \label{f0}
\ee

\begin{figure}
\begin{center}
\leavevmode
\epsfysize=300pt{\epsffile{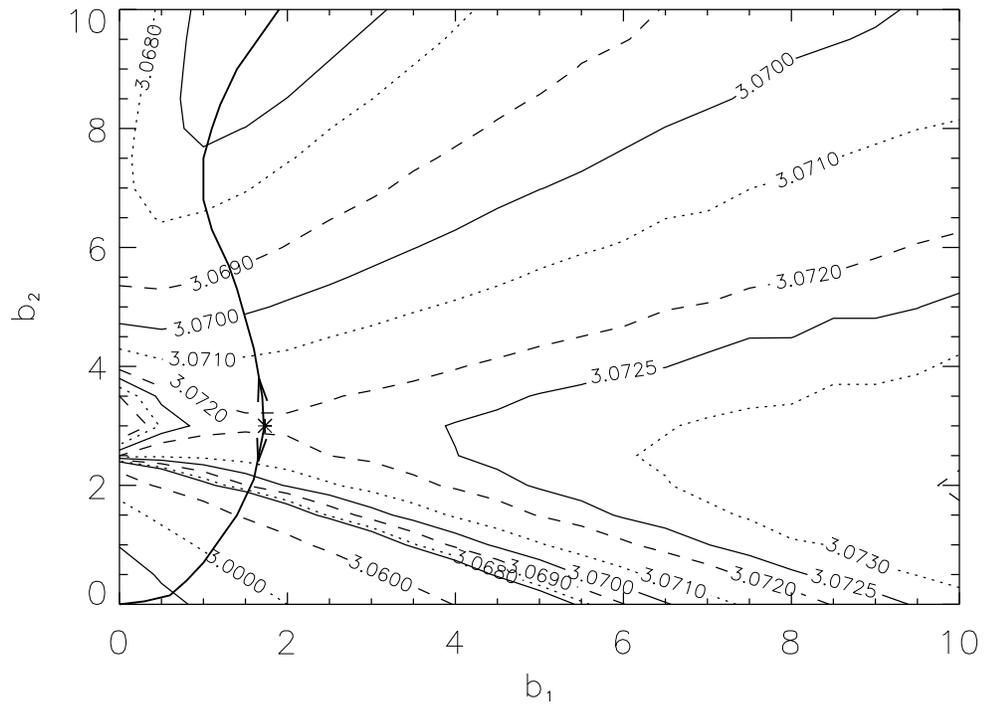}}
\end{center}
\protect\caption[5]{\label{fig:contour}Contour plot of the effective
central charge $c(6,8)$ along the self-dual surface $(b_1,b_2)$ for four
coupled three-state Potts models. The decoupled fixed point is shown as an
asterisk, and renormalisation group flow lines are sketched as a guide to
the eye.}
\end{figure}

A contour plot of $c(6,8)$, based on a grid of $21 \times 21$
parameter values for $(b_1,b_2)$, is shown in Fig.~\ref{fig:contour}.
The data for $c(4,6)$ look qualitatively similar, but are
less accurate due to finite-size effects. We should
stress that even though the absolute values of $c(6,8)$
are some 4 \% below what one would expect in the
$L \to \infty$ limit, the {\em variations} in $c$ are
supposed to be reproduced much more accurately \cite{Dotsenko}.
On the figure $q=3$, but other values of $q$ in the range $2 < q \le 4$
lead to similar results.

According to Zamolodchikov's
$c$-theorem \cite{Zamolodchikov}, a system initially in the vicinity of
the decoupled fixed point $(b_1,b_2) = (\sqrt{q},q)$,
shown as an asterisk on the figure,
will start flowing downhill in this central charge landscape.
Fig.~\ref{fig:contour} very clearly indicates that the decoupled fixed
point acts as a saddle point, and there are thus only two possibilities
for the direction of the initial flow.

The first of these will take the
system to the stable fixed point at the origin which trivially corresponds
to one selfdual $q^4$-state Potts model. For $q=3$ this leads to the
generation of a finite correlation length, consistent with
$c_{\rm eff} = 0$ in the limit of an infinitely large system.
As expected, the flow starts out in the $b_2$ direction, meaning that
it is the energy-energy coupling between layers ($K_2$) rather than
the spin-spin coupling within each layer ($K_1$) that controls the
initial flow.

More interestingly, if the system is started out in the {\em opposite}
dirrection (i.e., with $K_2$ slightly positive) it will flow towards
a third non-trivial fixed point,
for which the edge weights tend to infinity in some definite ratios.
[Exactly what these ratios are is difficult to estimate, given that the
asymptotic flow direction exhibits finite-size effects.]
Seemingly, at this point the central charge is only slightly lower than at
the decoupled fixed point, as predicted by the perturbative
renormalisation group \cite{Dotsenko}. From the numerical data we would
estimate the drop in the central charge as roughly
$\Delta c = 0.01$ -- $0.02$, in good agreement with the perturbative
treatment which predicts $\Delta c = 0.0168 + {\cal O}(\varepsilon^5)$
\cite{Dotsenko}.

All of these facts are in agreement with the conjectures put
forward in Ref.~\cite{Dotsenko}, and in particular one would think
that this third fixed point corresponds to a conformal field theory
with a non-abelian extended $S_4$ symmetry.

Finally, the numerics for $q=2$ (four coupled Ising models) is less
conclusive, and we cannot rule out the possibility of a more involved
fixed point structure. In particular, a $c=2$ theory is not only
obtainable by decoupling the four models, but also by a pairwise coupling
into two mutually decoupled four-state Potts (or Ashkin-Teller) models.
Indeed, a similar phenomenon
has already been observed for the case of {\em three} coupled Ising models
\cite{Dotsenko}.

\vspace{0.5cm}

\noindent{\large\bf Acknowledgments}

The author is indebted to M.~Picco for some very useful discussions.

\end{document}